# Charge instability of topological Fermi arcs in chiral crystal CoSi

Zhicheng Rao[1,2,†], Quanxin Hu[1,2,†], Shangjie Tian[3,†], Shunye Gao[1,2], Zhenyu Yuan[1,2], Cenyao Tang[1,2], Wenhui Fan[1,2], Jierui Huang[1,2], Yaobo Huang[4], Li Wang[5], Lu Zhang[1,2], Fangsen Li[5], Huaixin Yang[1,2,6], Hongming Weng[1,2,6], Tian Qian[1,2,6], Jinpeng Xu[1,2,7*], Kun Jiang[1,2], Hechang Lei[3*], Yu-Jie Sun[8,1*] and Hong Ding[1,6,7]

[1] *Beijing National Laboratory for Condensed Matter Physics and Institute of Physics, Chinese Academy of Sciences, Beijing 100190, China*

[2] *School of Physical Sciences, University of Chinese Academy of Sciences, Beijing 100049, China*

[3] *Department of Physics and Beijing Key Laboratory of Opto-electronic Functional Materials & Micro-nano Devices, Renmin University of China, Beijing 100872, China*

[4] *Shanghai Synchrotron Radiation Facility, Shanghai Institute of Applied Physics, Chinese Academy of Sciences, Shanghai 201204, China*

[5] *Vacuum Interconnected Nanotech Workstation (Nano-X), Suzhou Institute of Nano-Tech and Nano-Bionics (SINANO), Chinese Academy of Sciences (CAS), Suzhou 215123, China*

[6] *Songshan Lake Materials Laboratory, Dongguan, Guangdong 523808, China*

[7] *CAS Center for Excellence in Topological Quantum Computation, University of Chinese Academy of Sciences, Beijing 100190, China*

[8] *Department of Physics, Southern University of Science and Technology, Shenzhen 518055, China*

†These authors contributed equally to this work

*Correspondence to: sunyj@sustech.edu.cn, hlei@ruc.edu.cn, xujp@iphy.ac.cn

**Topological boundary states, emerged at the spatial boundary between topological non-trivial and trivial phases, are usually gapless, or commonly referred as metallic states. For example, the surface state of a topological insulator is a gapless Dirac state[1-3]. These metallic topological boundary states are typically well described by non-interacting fermions[1-5]. However, the behavior of topological boundary states with significant electron-electron interactions, which could turn the gapless boundary states into gapped ordered states, e.g., density wave states or superconducting states, is of great interest theoretically, but is still lacking evidence experimentally[14-18]. Here, we report the observation of incommensurable charge density wave (CDW) formed on the topological boundary states driven by the electron-electron interactions on the (001) surface of CoSi. The wavevector of CDW varies as the temperature changes, which coincides with the evolution of topological surface Fermi arcs with temperature. The orientation of CDW phase is determined by the chirality of the Fermi arcs, which indicates direct association between CDW and Fermi arcs. Our finding will stimulate the search of more interactions-driven ordered states, such as superconductivity and magnetism, on the boundaries of topological materials.**

The discovery of topological materials revolutionized our view on phase classification in condensed matter physics[1-5]. A common characteristic of topological materials, like topological insulators and semimetals, is the presence of nontrivial gapless boundary states, which are normally robust against symmetry-allowed perturbations owing to topological protection of the bulk. The topological boundary states not only directly reflect their topological properties, but also result in many exotic physical phenomena, such as quantum Hall conductivity[6,7], chiral magnetic effect[8] and anomalous quantum oscillations[9-10]. Typically, the physics of topological boundary states of topological materials are well captured from the perspective of non-interacting fermions. For example, density functional calculations have scored tremendous successes in predicting of topological electronic materials and their boundary states[11-13]. On the other hands, electron-electron interactions also give rise to many fascinating phenomena, including unconventional superconductivity[14, 15], Mott transition[16] and charge density wave[17,18]. Therefore, the destiny of topological boundary states under electron-electron interactions emerges as an interesting question.

In Dirac and triply-degenerate semimetals, the "Fermi arcs" formed by topological boundary states are not topologically protected since the Chern number is zero for Dirac nodes and undefined for three-degenerate nodes[19-22]. Though the topological boundary states form robust Fermi arcs in Weyl semimetals, such as TaAs-family[23,24], (Mo,W)Te$_2$[25,26] and Co$_3$Sn$_2$S$_2$[27], all these Weyl semimetals are not good candidates for studying the many-body effect of topological boundary states because of narrow nontrivial energy windows, short Fermi arcs, and coexisting other trivial states. Drumhead-like boundary states in topological nodal-line semimetals have been proposed as an ideal candidate for the many-body effect in theory due to the nearly flat surface band with large density of states in the vicinity of Fermi energy ($E_F$)[28], However, experimental evidence is still elusive.

Recently, transition metal silicides (CoSi-family) with nonsymmorphic space group $P2_13$, dubbed as chiral crystals, have attracted wide attentions[29-35]. Since the topological nodes of bulk states at the Brillouin zone (BZ) center Γ and corner R points carry opposite Chern numbers near $E_F$[31-34]. On the surface of chiral crystals, the Fermi surface (FS) just consists of two Fermi arcs traversing the whole surface BZ (Fig. 1b). The extremely long Fermi arcs and its simplicity of the chiral crystals provide an ideal platform to explore the many-body effect induced by the topological boundary states. In this work, we observe an incommensurable charge density wave driven by electron-electron interactions of the Fermi arcs on the (001) surface of CoSi.

We carried out scanning tunneling microscopy/spectroscopy (STM/S) experiments on the atomically flat CoSi surfaces prepared by *in-situ* repeated sputtering and annealing (see Method for details). Figure 1e presents a STM topographic image obtained on the (001) surface at 4.2 $K$. The square lattice with a period of ~4.43Å (the lattice constant is 4.44 Å) shows that the structure of the (001) surface is stable after the sputtering-annealing process. In addition, a stripe-like superlattice modulation is also discernible. We extracted the period $\lambda$ and angle $\alpha$ of the superlattice modulation by fast Fourier transformation (FFT), as shown in Fig. 1g. The value of $\lambda$ is about 8.16 $a_0$ and the angle $\alpha$ between the superlattice and the [100] direction is about 34.2°, which indicates that the stripe-like superlattice modulation has an incommensurable period. We then performed STS measurements on the stripe superlattice region (Fig. 1c), and found that the differential conductance spectrum exhibits an energy gap-like feature extending from -22meV to +20meV around $E_F$, which may be the insulating gap induced by the stripe superlattice.

We also performed STM measurements on the (111) surface of CoSi, on which the Fermi arcs are absent because the $\Gamma$ and R points are projected onto the same point on surface BZ. Figures 1f,h show the atomically resolved hexagon lattice while the stripe superlattice modulation disappears on the surface. The fact that the stripe superlattice only exists on a specific surface implies that its origin is not from the bulk states. In addition, we carried out high-resolution transmission electron microscope (TEM) measurements on the (001) surface of CoSi (Fig. 1d). The absence of the satellite spots corresponding to the stripe-like superlattice in the selected area electron diffraction (SAED) pattern, which also confirms that it is absent in the bulk of CoSi. Therefore, we conclude that the stripe-like superlattice originates from the (001) surface of CoSi.

We next demonstrate that evolution of superlattice modulation with temperature in Fig. 2 and Extended Data Fig. 3. A similar stripe-like superlattice clearly appears at 77 K but becomes less obvious at 300 K. The enlarged FFT images corresponding to the STM topographic data at different temperatures are plotted in Figs. 2d-f, showing that the spots of superlattice are well-defined at low temperatures but fade away at 300 K. Figure 2g shows that the peak of the spots in the FFT image broadens as the temperature is raised, and the corresponding peak becomes ill-defined at the room temperature, suggesting that the period of modulation is gradually melting (Fig. 2g). We summarize the spots of superlattice in Fig. 2h and Extended Data Fig. 3a. Those spots are distributed regularly in the Fourier space with both the length and angle of the spots decreasing with increasing temperature. It is known that there are several possibilities that can induce the stripe-like superlattice modulation on the surface, such as surface reconstruction, Moiré pattern, surface strain and charge density wave (CDW). The possibilities of surface reconstruction, Moiré pattern and surface strain can be ruled out (see Method for details). We thus conclude that the stripe-like superlattice modulation is a CDW order phase on the (001) surface.

In order to explore the reason why the wave vector of CDW varies with temperature, we measured the Fermi surface at different temperatures by angle-resolved photoemission spectroscopy (ARPES). Figures 3a,b show ARPES intensity maps at $E_F$ at 12 K and 195 K, respectively. The FSs at different temperatures consist of two similar Fermi arcs connecting $\bar{\Gamma}$ and $\bar{M}$. However, it is noteworthy that the Fermi arcs undergo significant deformation at different temperatures. Comparing the momentum distribution curve (MDC) at $E_F$ of the spectrum along the cut indicated in Fig. 3a at different temperatures, we found that the momentum positions of the Fermi arcs have obvious shift. The peak position of the MDC at $E_F$ at 12 K is the same as the one of the MDC at 18 meV below $E_F$ at 195 K, indicating that the surface chemical potential shifts up 18 meV with increasing temperature (Fig. 3f). Since the Fermi arcs are formed by the helical surface states[36], the lifting of the surface chemical potential results in the two Fermi arcs approaching to each other (see Method for details). We extracted the momentum locations of the Fermi arcs at different temperatures by tracking the peak positions of spectrum and plotted them together in Fig. 3c. We found that the two Fermi arcs change significantly at different temperatures. So, the observed temperature dependence of the wave vector of CDW is understandable.

We notice that the two long Fermi arcs near the time-reversal invariant point $\bar{X}$ are almost parallel, which is in accord with the FS nesting condition. When the Fermi arcs approach to each other with increasing temperature, the length and angle of nesting condition gradually decrease. The temperature evolution of Fermi arcs is consistent with the variation of wave vector of CDW, suggesting that the CDW is related to the Fermi arcs on the surface. Furthermore, we plotted the extracted $\vec{q}_{CDW}$ at 4.2 K on FS in Fig. 3a ($\vec{q}_{CDW}$ at 77 K and 300 K on FS in ExtendedData Fig.4). The good agreement between the nesting condition of two Fermi arcs and the wave vector of CDW provides a possible explanation that the CDW is induced by the partially nested Fermi arcs, which originates from the electron-electron interactions.

In CoSi crystals, there are two kinds of enantiomers with opposite chirality in the lattices (Extended Data Fig. 1a,b), while the bulk band dispersions are exactly the same, but the signs of Chern numbers at the topological nodes are reversed since the reversal of chiral lattice, resulting in the reversed surface Fermi arcs on the (001) surface (Extended Data Fig. 1, details in Method). If the CDW is directly related to the Fermi arcs, it should be mirrored on these two enantiomers of CoSi. To verify it, we carried out measurements on the two enantiomers of CoSi. Firstly, we performed low energy electron diffraction (LEED) measurements on the (001) surface of CoSi to distinguish the two enantiomers, and the mirrored Z-shaped enhanced intensity distribution manifests the opposite chirality between the two enantiomers (Figs. 4a, b). In Figs. 4c, d, the ARPES intensity maps on the two enantiomers show the mirrored configurations of the Fermi arcs, which indicate the chirality of Fermi arcs reversals (Extended

Data Fig. 1). Secondly, we observed mirrored stripe superlattices on the two enantiomers (Figs. 4e, f). This mirror effect can be also seen from the FFT of topographic image (Figs. 4e, f). The mirrored CDW verifies that it is related to the nesting condition of Fermi arcs. In addition, the two enantiomers of CoSi host exactly the same bulk FS and the same trivial pockets at the $\bar{X}$ point (Figs. 4c, d). Consequently, it furtherly indicates that the CDW is directly related to the surface Fermi arcs.

Our results provide strong evidence for the charge instability related to topological surface Fermi arcs towards a CDW phase in chiral crystal CoSi, which is the first experimental observation of many-body effect on topological boundary states. Unlike the conventional CDW phase[37-40], the wavevector of this kind of CDW is temperature-dependent. As a result, it is easy to tune the CDW phase continuously by temperature. The interaction-driven charge instability of surface Fermi arcs represents a new avenue for investigating the interplay between many-body interactions and topological boundary states. Whether other many-body responses, such as unconventional superconductivity and long-range magnetic orders, can be linked to topological boundary states remain to be further explored.

**Acknowledgements**

We thank Lu Yu, Chen Fang, Tiantian Zhang and Liqing Zhou for valuable discussions. We thank for technical support from Nano-X from Suzhou Institute of Nano-Tech and Nano-Bionics, Chinese Academy of Sciences (SINANO). This work was supported by the National Natural Science Foundation of China (U1832202, 11888101, 11920101005, 11822412, 11774423), the Chinese Academy of Sciences (QYZDB-SSW-SLH043, XDB33020100, XDB28000000), the Beijing Municipal Science and Technology Commission (No. Z171100002017018, Z200005), National Key R&D Program of China (No. 2018YFE0202600), the Fundamental Research Funds for the Central Universities and

Research Funds of Renmin University of China (RUC) (Grant No. 18XNLG14, 19XNLG13, 19XNLG17, 20XNH062), Beijing National Laboratory for Condensed Matter Physics.

**Data and materials availability:** All data needed to evaluate the conclusions in the paper are present in the paper and/or the Supplementary Materials. Materials and additional data related to this paper may be requested from the authors.

**Methods**

**Materials:** Single crystals of CoSi were grown by the chemical vapor transport method. Co and Si powders in 1:1 molar ratio were put into a silica tube with a length of 200 mm and an inner diameter of 14 mm. Then, 200mg $I_2$ was added into tube as a transport reagent. The tube was evacuated down to $10^{-2}$ Pa and sealed under vacuum. The tubes were placed in a two-zone horizontal tube furnace, the source and growth zones were raised to 950 K and 800 K in 2 days, and were then held there for 7 days. Shiny crystals with lateral dimensions up to several millimeters can be obtained.

**Surface preparation of CoSi:** The structure of CoSi single crystal is sample cubic. It is difficult to obtain a flat surface by mechanical cleaving owing to the strong covalent bond and lack of a preferred cleaving plane. So, we used the *in-situ* sputtering and annealing method to obtain an atomically flat plane on the (001) and (111) surfaces[41]. Firstly, we verified the (001) and (111) surfaces of CoSi by Laue and XRD and subsequently mechanically polished to a mirror finish on the surface. The atomically flat surface was prepared by Ar ion sputtering for

1h and annealing at 760℃ for 30 minutes repeatedly in an ultra-high vacuum with a base pressure of $1×10^{-9}$ $torr$. Reflection high-energy electron diffraction (RHEED) was carried out to monitor the quality of sample surface.

**STM and ARPES measurements:** STM Experiments were carried out in a low temperature ultrahigh vacuum STM system, Unisoku USM-1300. Topographic images were acquired in the constant-current mode with a tungsten tip. Before measurements, STM tips were heated by e-beam and calibrated on a clean Ag surface. Differential conductance spectra were acquired by a standard lock-in technique at a reference frequency of 973 Hz. ARPES measurements were performed at the 'Dreamline' beamline of the Shanghai Synchrotron Radiation Facility (SSRF) with a Scienta Omicron DA30L analyzer. The Fermi level is calibrated by measuring a clean gold sample at corresponding temperatures. All measurements were *in-situ* preformed.

**Transmission electron microscopy measurements:** The microscopic characterizations of CoSi were performed with a JEOL-2100F electron microscope operated at a voltage of 200 kV. The specimen was prepared by crashing the crystals in acetone and deposited them on a carbon film suspended on a copper grid. *In-situ* low-temperature TEM experiments were performed using a liquid-nitrogen-cooled specimen holder. High quality of CoSi crystals was confirmed by HRTEM.

**Alternative explanations of stripe-like superlattice on CoSi:**

1. Surface reconstruction: Surface reconstruction refers to the process by which atoms at the surface of a crystal form a different structure than that of the bulk. The atoms at or near the surface only experience inter-atomic interactions from one direction. The imbalance of chemical environment on the surface results in the positions of them with different spacing and/or symmetry from that of the bulk atoms, creating different surface structure and/or different positions of surface atoms relative to the bulk positions.

In the (001) surface of CoSi, we observed a square lattice with the lattice constant of 4.43Å (check the number in fig.1), consistent with the bulk symmetry, which is very close to the bulk lattice constant of 4.44Å measured by Laue diffraction. The same symmetry and lattice period can rule out the type of surface structure with different symmetry from the bulk, like the herring bone reconstruction on Au (001) surface. As for the surface reconstruction where surface atoms change positions and spacing, since the surface structure is a square lattice that is nearly the same as the bulk, there are two situations: one is that the direction of the surface square lattice is along the bulk lattice. This type of reconstruction can form a stripe pattern when one side of the surface square lattice shortens slightly and becomes a rectangle lattice. But the stripe pattern formed by this situation is along the lattice direction, which is contrary to the observed incommensurable stripe-like superlattice. The other type is that the surface lattice is rotated by an angle relative to the bulk lattice. This type reconstruction cannot form the stripe pattern. Combining the above information, we can rule out the possibility of surface reconstruction on the (001) surface of CoSi.

2. Moiré pattern: Moiré pattern is an interference pattern produced by overlaying a similar but slightly offset or rotated pattern. The Moiré phenomenon is commonly discussed for two-dimensional materials. The effect occurs when the lattice constants or/and angle of a 2D layer mismatch with those of the substrate. In CoSi single crystal, the structure of the underlying bulk is a square lattice with the period of 4.44 Å and the surface lattice is also a square structure with the almost same period. The static stripe-like pattern cannot be created by superimposing two square grids by offset and rotation. In fact, this static stripe-like Moiré pattern can be obtained by superimposing a square grid and a rectangle grid (the long side of the rectangle is equal to the side of the square), but the stripe-like moiré patter is along the side of grids, which is not consistent with the observed incommensurable stripe on the CoSi surface. So, the stripe-like superlattice cannot be a Moiré pattern. Furthermore, in Figs. 2a-c, the lattice direction on the surface remains almost in the same direction but the angle $\alpha$ of the stripe-like modulation varies from 34.2° to 13.5° at different temperatures, which can also rule out the possibility of Moiré pattern.

3. Surface strain: The stripe pattern can also be obtained when the uniaxial strain is applied. On the Au (111) surface, the nature of its stripe pattern is the surface reconstruction induced by the uniaxial strain. On the other hand, the epitaxial strain induces deformation of the surface lattice, resulting in a stripe pattern. The nature of this type stripe pattern is also Moiré pattern. So, we ruled out this possibility.

**Effect of defects and/or impurities for the CDW:** In addition to the experimental results mentioned in the main text, we have also performed measurements on different samples. Figure S3a shows the distribution of charge density wave in the Fourier space. At same temperature, we observed that the wave vector is not exactly the same at different regions, even the positions of some spots have some differences, but they remain clustering together. We have noted that many defects and/or impurities are randomly distributed on the (001) surface from the topographic data. Energy-dispersive X-ray spectroscopy measurements also show the ratio of Co and Si is not strictly 1:1 (about 0.86-0.97:1). When these defects are intensively distributed in some regions, the charge density wave can even be destroyed, as shown in Fig. 1e. We propose that the presence of defects/impurities can significantly affect their vicinity in two ways: First, the random distribution of defects/impurities may cause the local chemical potential shift. Like the temperature can shift the surface chemical potential, defects can also change the Fermi surface by shifting the local chemical potential, resulting in variation of the wave vector of CDW. Second, defects may scatter the CDW, resulting in the distribution of CDW deviating slightly from the trend line. In Fig.S3b, we summarize the relationship between the length and the angle of the wave vector of CDW. Although the distributions of some spots have some overlap at different temperatures, the center of distributions at different temperatures shows that the length and the angle of $q_{CDW}$ decrease with rising temperature.

**Helical surface states in chiral crystal:** Since the bulk band structure has three- and four-degenerate nodes at Γ and R, respectively, which carry nonzero Chern numbers ±2. On the (001) surface, the two topological nodes are projected to the surface BZ center $\bar{\Gamma}$ and corner $\bar{M}$. Those projections of topological nodes are surrounded by the helical surface states, which are topological equivalent to the non-compact Riemann surface. The helical surface states form open Fermi arcs at constant-energy contours, which connect $\bar{\Gamma}$ and $\bar{M}$. With lifting the constant energy, the Fermi arcs rotate clockwise and anticlockwise with respect to $\bar{\Gamma}$ and $\bar{M}$, respectively. Because two ends of Fermi arcs are pined at $\bar{\Gamma}$ and $\bar{M}$, the two Fermi arcs around $\bar{X}$ (as shown in Extended Data Fig. 4) will approach gradually and touch each other with energy shift.

**Wavevector of CDW on ARPES constant-energy contours:** The averaged wavevector of CDW at different temperatures are plotted on the ARPES constant-energy contours. We find that the wavevectors of the CDWs at 77 K and 300 K coincide with the nesting condition of two Fermi arcs on the constant-energy contours at 5 meV and 25 meV (the Fermi level at 12K is defined as 0), as shown in Extended Data Fig. 4. Considering that the surface chemical potential is shifted up about 18 meV when temperature is raised to 190 K, the energy positions of the two constant-energy contours are reasonable. Actually, the Fermi arcs around the time-reversal invariant point $\bar{X}$ is the only one region on the FSs of (001) surface that matches the wavevector of CDW in both magnitude and trend. The presence of charge instability typically removes the nested portions of the Fermi surface. The ARPES intensity maps in Extended Data Fig.4 are measured with left- and right-circularly polarized light. We note that the intensity of Fermi arcs around the $\bar{X}$ point is still missing, as shown as the blue arrow in Extended Data Fig. 4. Due to the uneven distribution of CDW in the real space and the limitation of energy resolution, we cannot be sure that this missing intensity of Fermi arcs is induced by the charge instability (the matrix element effect can also induce a similar missing of intensity[42]), which requires further research.

**Reversal helical surface states on two kinds of enantiomers of CoSi**

Both of two enantiomers of CoSi crystalize in the same space group ($P_{2_13}$), which have no mirror, inversion and roto-inversion symmetries. In crystal structure of two enantiomers, Co and Si atoms have exchanged positions with each other (ExtenedData Fig. 1a,b). From the view of [111] direction, the helical arrangement of Co and Si atoms reverse their chirality in the two enantiomers. Since two enantiomers belong to the same space group, bulk band dispersions of them are exactly the same (ExtendedData Fig. 1d,e). As we know, the surface projection of a topological nodes with nonzero Chern number is surrounded by helical surface states[36]. These surface states along the loop encircled the projection have chiral band dispersions. Consider a loop encircled the $\bar{\Gamma}$ point, as shown in Extended Data Fig. 1c, we observed two surface band dispersions with same chirality along the loop, but chirality of surface bands on the two enantiomers is opposite (ExtendedData Fig. 1f,g). These results provide sufficient evidence that the chirality of helical surface states is reversal around $\bar{\Gamma}$ point in two enantiomers, which indicate the topological nodes at $\Gamma$ point have opposite Chern number in two enantiomers.

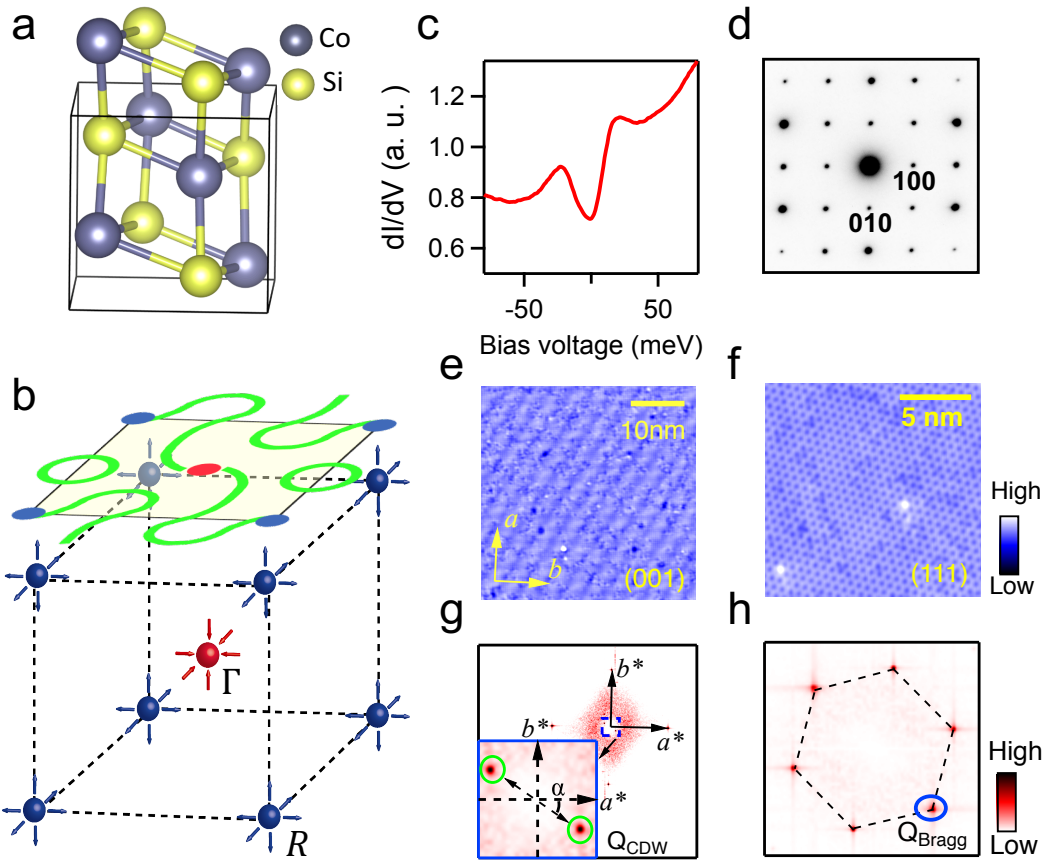

**Figure 1 | Stripe-like superlattice modulation on (001) surface of CoSi. a**, Crystal structure of CoSi. **b**, Schematic showing the bulk-boundary correspondence in CoSi, the topological nodes at the Γ and R points host nonzero Chern number ±2 and their projections on the (001) surface are connected by two Fermi arcs. **c**, Tunneling differential conductance spectrum, an energy gap-like feature extends from -20meV to +22meV around $E_F$. **d**, Typical selected area electron diffraction pattern taken along the [001] zone axis at 100 K, showing the stripe-like modulation absent in the bulk of CoSi. **e**, Atomically resolved STM topographic image (40nm × 40nm) of the CoSi (001) surface taken at 4.2K showing the stripe-like superlattice. **f**, STM topographic of the CoSi (111) surface taken at 4.2K, showing a hexagonal lattice structure. **g**, Fourier transform of the (001) surface topographic image of CoSi, the zoom-in image (insert) marked by the blue box show the spots (green circle) corresponding to the stripe-like superlattice modulation. **h**, Fourier transform of the (111) surface topographic image, showing no superlattice on the (111) surface.

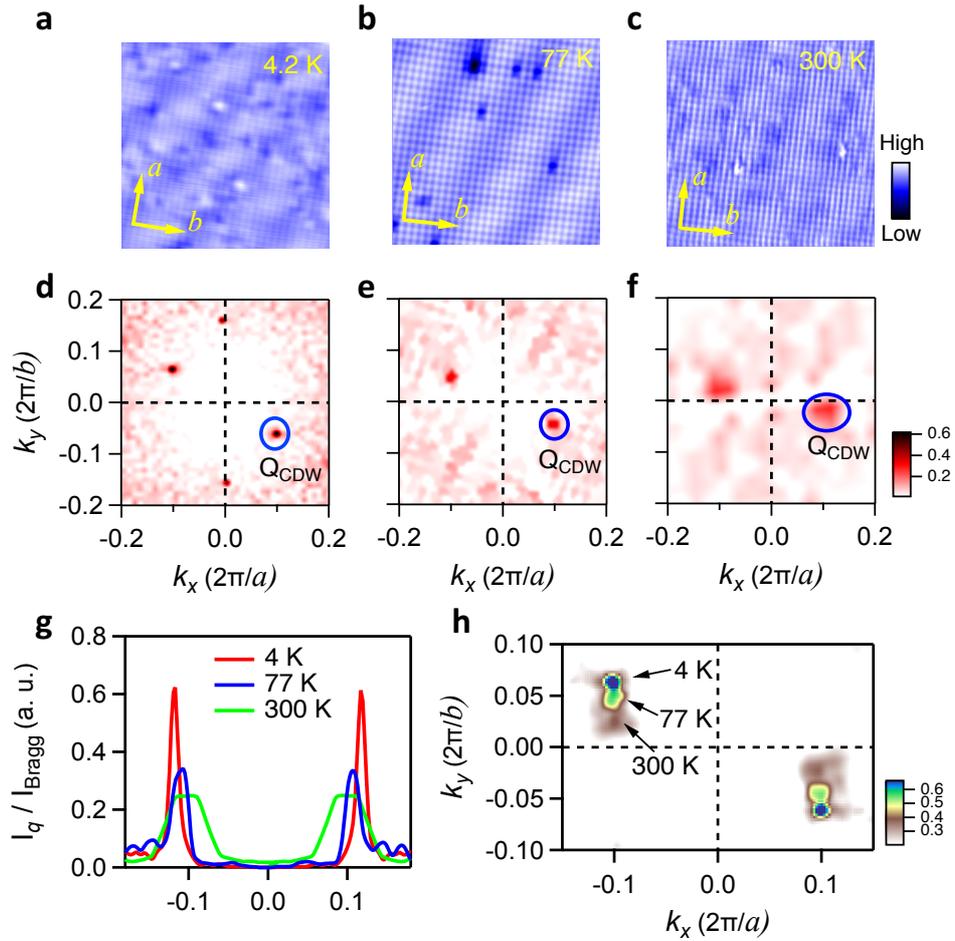

**Figure 2 | Evolution of stripe-like superlattice modulation with temperature. a-c,** STM topographic image (10nm × 10nm) of the (001) surface of CoSi showing the stripe-like superlattice at 4.2 K, 77 K and 300 K, respectively. **d-f,** Zoom-in Fourier transform images of **a-c,** with the spots corresponding to the stripe-like superlattice modulation marked by the blue circle. **g,** Intensity plot along two spots of the stripe-like superlattice in **d-e**, showing the period of superlattice slightly enlarging with temperature. **h,** The distribution of spots corresponding to the superlattice in the Fourier space at different temperatures, showing the superlattice varying with temperature.

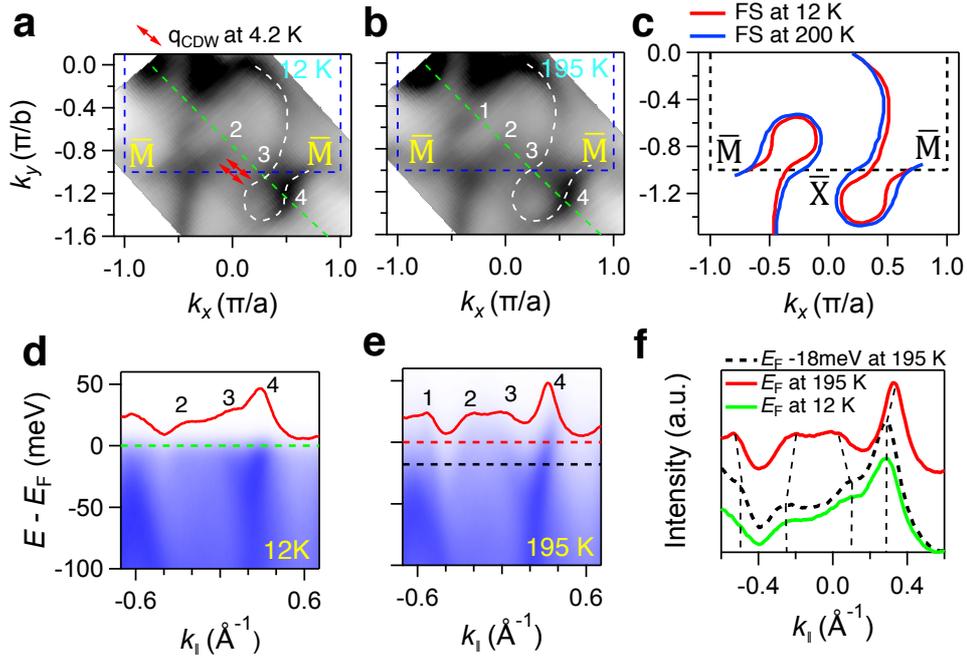

**Figure 3 | Surface chemical potential varing with temperature. a,b**, ARPES intensity maps at $E_F$, measured with 55 eV light at 12 K (**a**) and 195 K (**b**). The white dashed lines are the guides to the eye for the Fermi arcs. **c,** Extracted Fermi arcs from **a** (red line) and **b** (blue line), the two Fermi arcs at 195 K approach to each other with respect to that at 12 K. **d,e,** Band dispersion along the cut indicated in **a**, showing the momentum location of Fermi arcs shifting with temperature. **f**, MDCs of spectrum, the energy position of that is indicated in **d** and **e**. The peak position of MDC at $E_F$ at 12 K is the same with that at $E_F - 18$ meV at 195 K, showing that the surface chemical potential is shifted up by 18 meV.

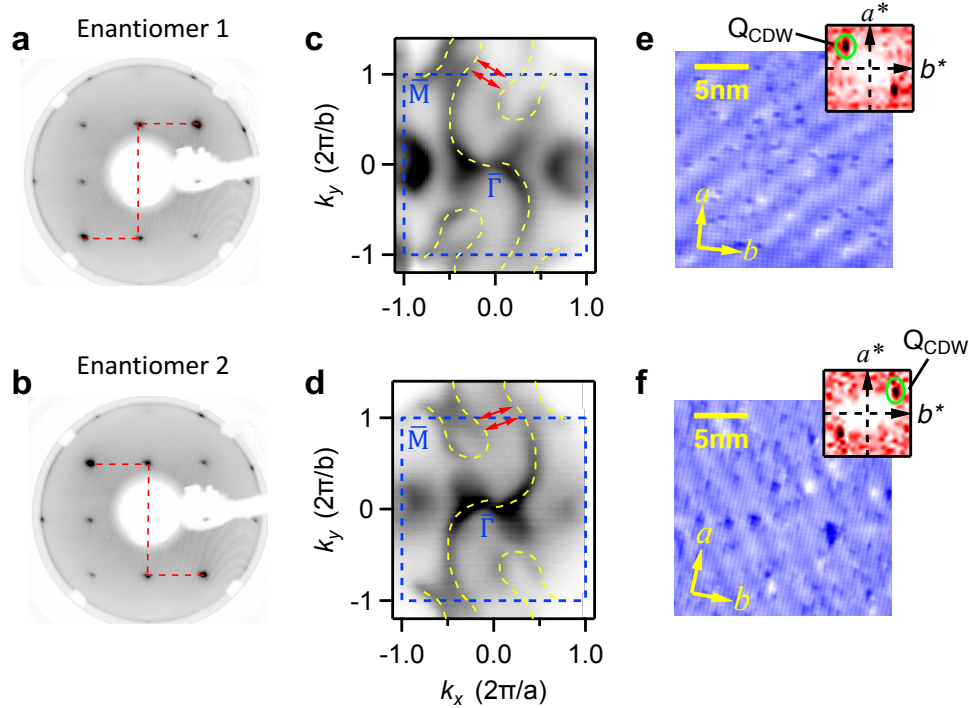

**Figure 4 | CDW phase on two enantiomers with opposite chirality. a,b**, LEED patterns of two enantiomers with opposite chirality, the mirrored Z-shaped (the red dashed lines as the guides to the eye) enhanced intensity diffraction spots show the opposite chirality of crystal structure of two enantiomers. **c,d**, ARPES intensity maps of two enantiomers at $E_F$, showing reversal chirality of Fermi arcs on the Fermi surface. The yellow dashed lines are the guides to the eye for the Fermi arcs. **e,f**, Atomically resolved STM topographic images (20 nm × 20 nm) of the two enantiomers taken at 4.2 K, showing different orientations of the CDW phase on two enantiomers. The spots (green circle) correspond to the CDW phase in the enlarge FFT image (insert), indicating that the CDW phases on two enantiomers of CoSi are mirrored.

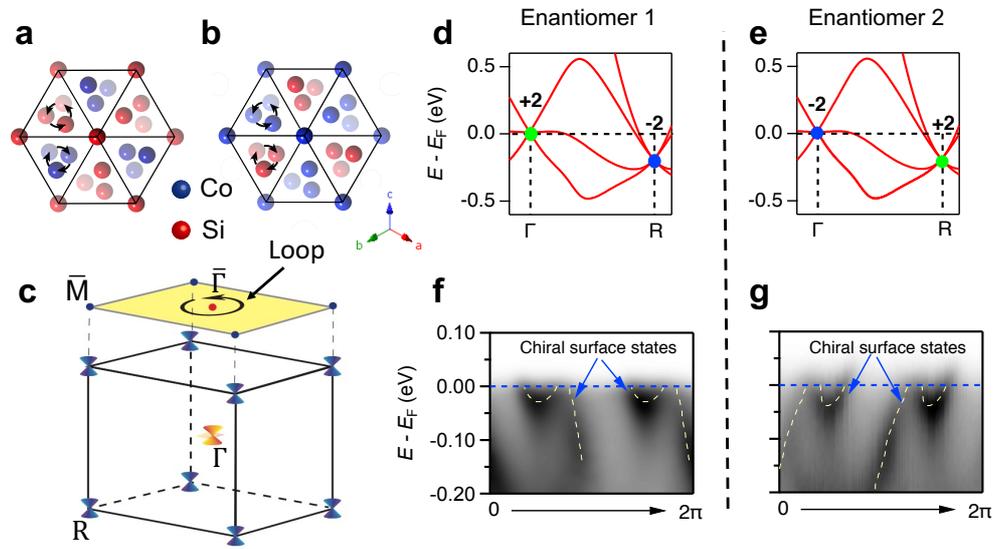

**ExtendedData Fig. 1 | Electronic structure in two enantiomers with opposite chirality. a,b,** Crystal structures of two enantiomers of CoSi in the view of [111] direction. The transparency of balls indicates the depth of the atomic position from top to bottom. c, Bulk BZ and (001) surface of CoSi. **d,e**, Calculated bulk band dispersions of two enantiomers of CoSi along $\Gamma - R$ direction,, the sign of Chern number at the $\Gamma$ and R points in two enantiomers is opposite. **f,g,** ARPES intensity plots showing the chirality of the surface band dispersion along the loop encircled the $\bar{\Gamma}$ point (indicted in **c**). The chirality of surface band dispersion on two enantiomers is determined by the sign of Chern number at the $\bar{\Gamma}$ point.

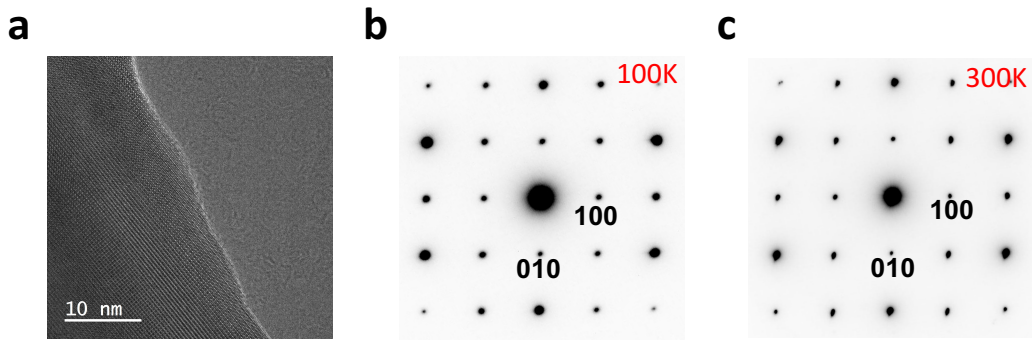

**Extended Data Fig. 2 | TEM results of CoSi along [001] direction. a,** High-resolution transmission electron microscopy (HRTEM) image of CoSi sample taken along the [100] zone-axis direction. **b,c,** Typical selected area electron diffraction (SAED) patterns taken along the [100] zone axis at 100 K and 300 K, respectively, showing the stripe-like superlattice absent in the bulk of CoSi.

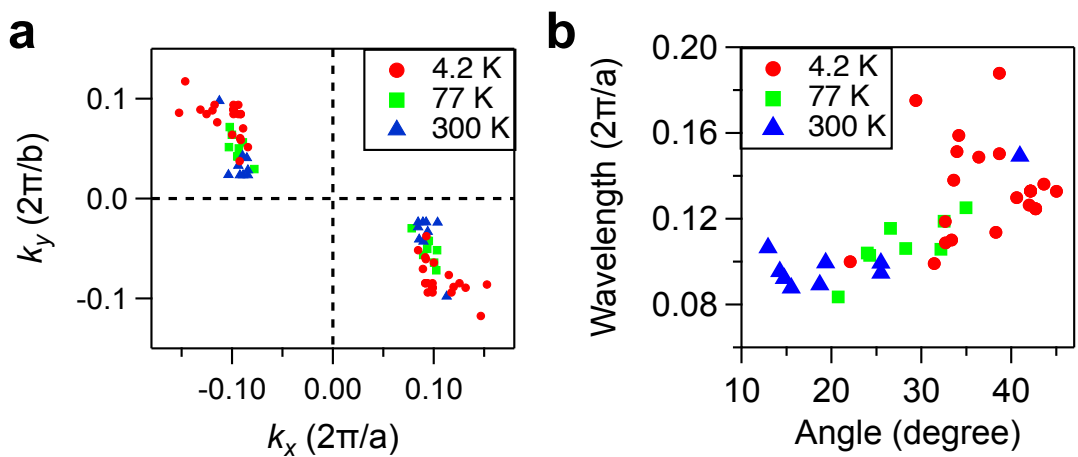

**Extended Data Fig. 3 | Distribution of charge order on different regions. a,** Spots

corresponding to the charge order on different regions at different temperatures in the Fourier space. **b**, Relationship between the length and the angle of the wave vectors at different temperatures.

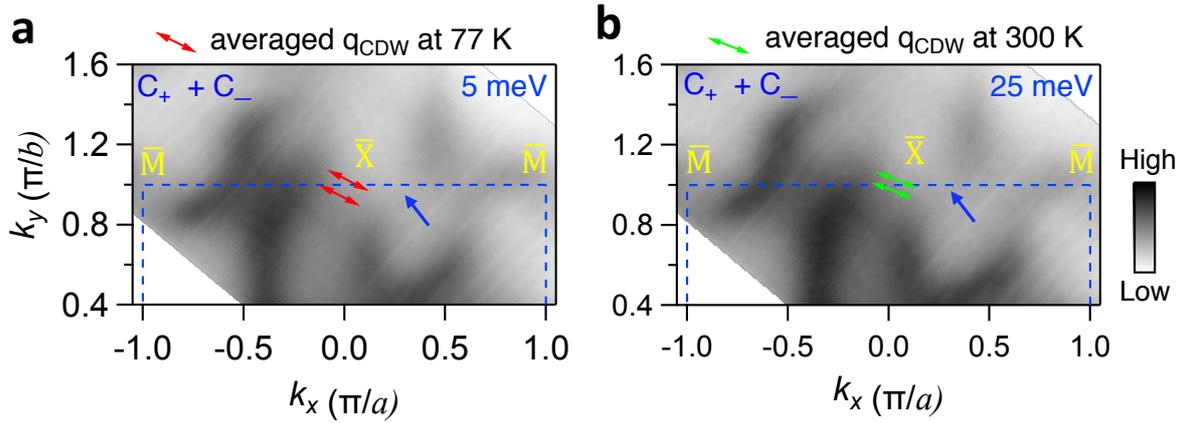

**Extended Data Fig. 4 | Relationship between the wave vector of CDW and the nesting condition of Fermi arcs at different temperatures. a,b**, Averaged wave vector of CDW at 77 K (**a**) and 300 K (**b**) on ARPES constant-energy contours at 5 meV (**a**) and 25 meV (**b**), respectively, showing a good consistence between the wave vector of CDW and the nesting condition of Fermi arcs. The value of $E_F$ at 12K is defined as 0. ARPES intensity maps are measured with left- and right-circularly polarized 55 eV light and the constant-energy contours is the sum of these two.